\documentstyle[12pt,titlepage]{article}
\begin{document}
\pagestyle{empty}
\newcommand{\lapproxeq}{\lower
.7ex\hbox{$\;\stackrel{\textstyle <}{\sim}\;$}}
\newcommand{\skips}{\vspace{0.2cm}}
\newcommand{\skipm}{\vspace{0.4cm}}
\baselineskip=0.212in

\begin{flushleft}
\large
{SAGA-HE-56-94  \hfill February 17, 1994}  \\
\end{flushleft}

\vspace{1.5cm}

\begin{center}

\Large{{\bf Nuclear Shadowing}} \\

\vspace{0.3cm}

\Large{{\bf in a Parton Recombination Model:~ Q$^2$ Variation }} \\

\vspace{1.2cm}

\Large
{S. Kumano $^\star$ }         \\

\vspace{1.0cm}

\Large
{Department of Physics} \\

\vspace{0.1cm}

\Large{Saga University}    \\

\vspace{0.1cm}

\Large
{Saga 840, Japan}         \\

\vspace{1.2cm}

\Large{ABSTRACT}

\end{center}

Q$^2$ variation of the nuclear-structure-function ratio
$F_2^A(x,Q^2)/F_2^D(x,Q^2)$
is investigated in a parton model with Q$^2$-rescaling
and parton-recombination effects.
Calculated results are compared with
the NMC (New Muon Collaboration) and
the Fermilab-E665 experimental data.
We find that our theoretical results show small Q$^2$ variations
and that they are consistent with the data
within present experimental accuracy.

\vspace{3.2cm}

\hfill

\vspace{0.5cm}

\vfill

\noindent
{\rule{6.cm}{0.1mm}} \\

\vspace{-0.4cm}

\noindent
\normalsize
{$\star$ ~Email: kumanos@himiko.cc.saga-u.ac.jp.} \\

\vspace{0.0cm}
\hfill
{submitted for publication}

\vfill\eject
\pagestyle{plain}


It is known that the structure functions F$_2$ for nuclei
are different from the one for the nucleon.
This is called the EMC (European Muon Collaboration) effect.
The differences are found originally at
medium and large Bjorken-$x$.
In recent years, many experimental data are obtained
in the small $x$ region, which is so called the shadowing region.
There are theoretical attempts to explain the EMC effect
in the medium-large $x$ region and the shadowing in the small $x$.
However, most models investigated both regions separately and
enough effort has not been
made for studying the structure function in the whole $x$ region.
On the other hand, it is urgent to know
nuclear parton distributions in the wide $x$ range in order
to use them for other topics
in high-energy heavy-ion physics.

It is not straightforward to study a model which is valid
in the wide $x$ region.
An attempt to combine the medium-$x$ physics and the small-$x$ one
in a dynamically consistent way is made in Refs. \cite{SK,OTHER}.
Our model is based on a parton model and is not on macroscopic
nuclear physics (nuclear binding, Fermi motion, vector-meson
dominance, and so on) explicitly.
We simply incorporated two mechanism in our model:
Q$^2$ rescaling and parton recombination.
The rescaling has been used for explaining the EMC effect
in the medium $x$ region and the recombination for
the shadowing in the small $x$.
We calculated nuclear parton distributions
by using these ideas at small Q$^2$ and obtained distributions
are evolved by using the Altarelli-Parisi equation \cite{SK}.
As a result, the ratio [$F_2^A(x)/F_2^D(x)>1$] in the large-$x$ region
is explained by quark-gluon recombinations.
The EMC effect in the medium-$x$ region is mainly due to
the Q$^2$-rescaling mechanism. The shadowing in the small-$x$ region
is due to modifications in gluon distributions.
Although our theoretical results depend on input sea-quark and
gluon distributions, we obtained reasonably good agreement with
the EMC, the NMC (New Muon Collaboration), and
the Fermilab-E665 experimental data
if we choose an appropriate set of parameters.
Although some Q$^2$ dependence is shown in Ref. \cite{SK},
detailed comparison with the NMC \cite{NMC}
and the Fermilab-E665 \cite{FNAL}
Q$^2$-variation data is not made.
The purpose of this paper is to supplement the previous
publication \cite{SK} by showing the Q$^2$ variation explicitly.


We calculate the Q$^2$ rescaling and the parton recombination at small
Q$^2$ ($\equiv$Q$_0$$^2$); then, the structure function
F$_2$ is evolved by using the Altarelli-Parisi equation.
We refer the reader to the paper in Ref. \cite{SK}
for a complete account of our model and formalism.
We briefly discuss our calculation in the following.
We first calculate nuclear parton distributions at Q$_0$$^2$
by using the Q$^2$ rescaling for valence quarks.
Sea-quark and gluon distributions are modified
so that the momentum conservation is satisfied.
Then, the obtained parton distributions are used for calculating
parton-recombination effects. In this way, we get parton distributions
with the Q$^2$-rescaling and parton-recombination effects at
Q$_0$$^2$. These distributions
are evolved by using the Altarelli-Parisi equation.
According to the results in Ref. \cite{SK} (Sec. IV-C), an appropriate
choice of Q$_0$$^2$ is 0.8 GeV$^2$ for explaining experimental
data. Parameters in our model are given in Sec. IV-B of Ref. \cite{SK}.
It should be noted that the Q$^2$ evolution in this research
is calculated by using the ordinary Altarelli-Parisi equation \cite{EVOL},
and modifications of the Q$^2$ evolution
due to parton recombinations \cite{QIU} are not taken
into account. To solve the integro-differential equation in Ref. \cite{QIU}
accurately is a significant research problem by itself, so we leave
the full Q$^2$ evolution issue as our future research topic.
Furthermore, the perturbative QCD would not be applicable
in the small-Q$^2$ region,
so our calculation should be considered as a naive estimate.
Our Q$^2$ evolution results are shown in Figs. 1, 2, and 3
together with the NMC \cite{NMC} and the E665 \cite{FNAL} data.

Figure 1 shows our results for the calcium nucleus
at $x$=0.0085, 0.035, 0.125, and 0.55.
$x$=0.0085 and 0.035 are in the shadowing region;
0.125 is in the anti-shadowing;
0.55 is in the region of the original EMC effect.
The solid [dashed] lines are the results by using
the MRS-1 (Martin-Roberts-Stirling set 1)
[the KMRS-B0 (Kwiecinski-Martin-Roberts-Stirling set B0)]
parametrization as input distributions. Our theoretical calculations
are compared with the NMC data in 1991 \cite{NMC}.
Used constants are the starting Q$^2$ (Q$_0$$^2$=0.8 GeV$^2$),
the cutoff for leak-out partons ($z_0$=2.0 fm), and
the rescaling parameters ($\xi_C^V$=1.60, $\xi_{Ca}^V$=1.86).
It is shown in Fig. 1 that the ratio $F_2^A(x,Q^2)/F_2^D(x,Q^2)$
slightly increases with $Q^2$ at small $x$ and that the ratio
is almost constant at medium $x$.
Considering the small Q$^2$ variations of our theoretical results
and the experimental errors, we find that our results are
consistent with existing experimental data.

Results in Fig. 2 are for the carbon nucleus.
It shows that our results agree with the experimental data
within the present experimental accuracy.
Calculated Q$^2$ variations are rather large at Q$^2\approx$1 GeV$^2$
in the small-$x$ region. In fact, the dotted curve (Q$^2$=0.8 GeV$^2$) in
Fig. 7b of Ref. \cite{SK} may seem contradictory to the experimental data.
However, the large Q$^2$ variation in Fig. 2 and the discrepancy from
the data in Fig. 7b of Ref. \cite{SK} should not be taken very seriously.
This is because the perturbative QCD, especially in the leading order,
would not work at small Q$^2$.
Furthermore, as shown in Figs. 2a and 2b,
the discrepancy is not so large and theoretical curves are
consistent with the data if we consider the experimental accuracy.

Next, our calculations are compared with the Fermilab-E665 experimental
data for the xenon nucleus \cite{FNAL} in Fig. 3.
The experimental data are taken from
the recent report in Ref. \cite{FNAL}
and they are in the $x$ range, $0.001<x<0.025$.
The theoretical curves are calculated at $x=0.01$ and $0.025$ in comparison.
The calculated ratio slightly increases with Q$^2$.
Considering that the E665 data at small Q$^2$ ($<$1 GeV$^2$) are obtained
at small $x$ ($\sim 0.002$) and that the data at larger Q$^2$ ($>$3 GeV$^2$)
are at $x$ ($\sim 0.02$), we find that
our theoretical results agree with the E665 data reasonably well.
{}From these comparisons in Figs. 1, 2, and 3,
we conclude that the Q$^2$ variations calculated
in our parton model are in agreement with the experimental data.
However, accurate experimental data are needed for testing
details of theoretical works. In the theory side, we need progress
in numerical analysis of the full nuclear Q$^2$ evolution and
in the next-to-leading-order effects \cite{KKK}.


Q$^2$ variations of the nuclear structure function
ratio $F_2^A(x,Q^2)/F_2^D(x,Q^2)$ are calculated
in a parton model. We incorporated two mechanisms:
the Q$^2$ rescaling and the parton recombination,
in our model in a dynamically consistent way.
Calculated results are compared with
the NMC and the Fermilab-E665 experimental data.
Our theoretical results show small Q$^2$ variation and
are consistent with existing experimental data.

$~~~$


S.K. thanks Drs. M. van der Heijden and C. Scholz for information
on the NMC data and Dr. H. Schellman for information on the E665 data.

\vfill\eject

\vspace{3.0cm}
\noindent
{\Large\bf{Figure Captions}} \\

\vspace{-0.38cm}
\begin{description}
   \item[Fig. 1]
Q$^2$ variations of the ratio $F_2^{Ca}(x,Q^2)/F_2^D(x,Q^2)$ at
(a) x=0.0085, (b) x=0.035, (c) x=0.125, and (d) x=0.55.
Experimental data are the NMC data (1991) in Ref. \cite{NMC}.
The solid (dashed) curves are the calculated results by using
the MRS-1 (KMRS-B0) distribution.
   \item[Fig. 2]
Q$^2$ variations of the ratio $F_2^{C}(x,Q^2)/F_2^D(x,Q^2)$.
See Fig.1 for the notations.
   \item[Fig. 3]
Q$^2$ variations of the ratio $F_2^{Xe}(x,Q^2)/F_2^D(x,Q^2)$.
The solid (dashed) curve is our result at $x$=0.025 (0.01)
by using the MRS-1.
Experimental data are the E665 data in the $x$ range
($0.001<x<0.025$) \cite{FNAL}.

\end{description}

\begin{thebibliography}{}
\bibitem{SK} S. Kumano, Phys. Rev. {\bf C48}, 2016 (1993).
\bibitem{OTHER} L. L. Frankfurt, S. Liuti, and M. I. Strikman,
                research in progress;
                S. Kulagin, G. Piller, and W. Weise,
                to be submitted for publication.
\bibitem{NMC}
P. Amaudruz et al. (NMC collaboration), Z. Phys. {\bf C51}, 387 (1991).
\bibitem{FNAL}
M. R. Adams et al. (Fermilab-E665 collaboration),
                   Phys. Lett. {\bf B287}, 375 (1992);
                   preprint FNAL-93/245, submitted to Phys. Rev. D.
\bibitem{EVOL} We used the Q$^2$ evolution subroutine in
S. Kumano and J. T. Londergan, Comput. Phys. Commun. {\bf 69}, 373 (1992).
\bibitem{QIU} A. H. Mueller and J. Qiu, Nucl. Phys. {\bf B268}, 427 (1986);
J. Qiu, {\it ibid.} {\bf B291}, 746 (1987).
\bibitem{KKK} R. Kobayashi, M. Konuma, and S. Kumano, research in progress.
\end{thebibliography}
\end{document}